\documentclass[journal,a4,comsoc]{IEEEtran}
\usepackage{graphicx}
\usepackage{amssymb}
\usepackage{amsmath}
\usepackage{cite}
\usepackage{mathrsfs}
\usepackage[displaymath,mathlines]{lineno}
\usepackage{color}
\usepackage{tabulary}
\usepackage{multirow}
\usepackage{pbox}
\usepackage{multicol}
\usepackage{lipsum}
\usepackage{amsthm}
\usepackage{relsize}
\usepackage{lipsum}
\usepackage{epstopdf}
\usepackage{float}
\usepackage{mathtools}
\usepackage{calc}
\usepackage{makecell}
\usepackage{caption}
\usepackage{subcaption}
\usepackage{algorithmic}
\usepackage{soul}
\usepackage{algorithm}
\makeatletter
\newcommand{\doublewidetilde}[1]{{%
		\mathpalette\double@widetilde{#1}}}
\newcommand{\double@widetilde}[2]{%
		\sbox\z@{$\m@th#1\widetilde{#2}$}%
		\ht\z@=.5\ht\z@
		\widetilde{\box\z@}}
\makeatother

\allowdisplaybreaks

\makeatother
\makeatletter

\usepackage{titlesec}

\begin{document}

\title{\huge Metaheuristic Optimization of Trajectory and Dynamic Time Splitting for UAV Communication Systems }
\author{Trinh Van Chien, Nguyen Minh Quan, Oh-Soon Shin, and Van-Dinh Nguyen  \vspace{-1cm}
\thanks{T. V. Chien and N. M. Quan are with the School of Information and Communications Technology, Hanoi University of Science and Technology, Hanoi 100000, Vietnam (e-mail: chientv@soict.hust.edu.vn and quan.nguyenminh1@hust.edu.vn). O.-S. Shin is with the School of Electronic Engineering, Soongsil University, Seoul 06978, South Korea (e-mail: osshin@ssu.ac.kr). V.-D. Nguyen is with College of Engineering and Computer Science and also with Center for Environmental Intelligence VinUniversity, Vinhomes Ocean Park, Hanoi 100000, Vietnam (e-mail: dinh.nv2@vinuni.edu.vn). This research is funded by the Vietnam Ministry of Education and Training under project number B2025-BKA-04. The work of Van-Dinh Nguyen is funded by the VinUniversity Seed Program.\textit{Corresponding author}: Van-Dinh Nguyen.}
}



\maketitle

\begin{abstract}
The integration of unmanned aerial vehicles (UAVs) into wireless communication systems has emerged as a transformative approach, promising cost-efficient connectivity. This paper addresses the optimization of the dynamic time-splitting ratio and flight trajectory for a communication system linking a ground base station to the UAV equipped with backscatter devices (referred to as UB), and from UB to an end user. Given the inherent non-convexity of the problem, we develop two meta-heuristic-based approaches inspired by genetic algorithm and particle swarm optimization to enhance the total achievable rate while reducing computational complexity. Numerical results demonstrate the effectiveness of these meta-heuristic solutions, showcasing significant improvements in the achievable rate and computation time compared to existing benchmarks.

\end{abstract}

 \vspace{-0.25cm}
\section{Introduction}
Unmanned aerial vehicles (UAVs) have recently witnessed a significant surge in deployment across various sectors, ranging from military operations to package delivery, search and rescue missions, data collection, and facilitation of wireless communications \cite{ghamari2022unmanned}. Upon integration into wireless networks, UAVs serve as versatile components, primarily functioning as aerial base stations (BSs). This integration enhances the quality of service and extends coverage to remote and challenging environments \cite{10098686}. Particularly noteworthy is their role as intermediaries during peak hours when direct communication links face obstructions or equipment failures at BSs  \cite{8918497}.
However, UAVs encounter limitations due to restricted energy capacity, dictated by their compact size and weight. This necessitates managing dual energy requirements, encompassing both data transmission and movement. Prioritizing energy efficiency is paramount to ensure prolonged functionality and uninterrupted operations. One effective approach to address these challenges is leveraging backscatter for low power consumption and implementing caching mechanisms \cite{9723521}.

Backscatter communication (BackCom) has garnered significant interest in UAV research due to its minimal power consumption attributes. Researchers have been exploring innovative protocols and resource management strategies to enhance system performance in UAV networks equipped with backscatter devices \cite{8901136}. These protocols aim to streamline scheduling, plan trajectories, and allocate transmit power effectively.
Despite its potential advantages, challenges such as the limited battery capacity of UAVs prompt further investigation into energy-saving solutions like wireless power transfer (WPT). Recent research has focused on various aspects of integrating UAVs into wireless networks, with particular attention to caching issues. For instance, Zhang \textit{et al.} employed latent Dirichlet allocation to analyze user requests and optimize caching strategies \cite{zhang2021intelligent}. The work in \cite{nasir2021latency} not only optimized UAV location, caching, and computing resources but also proposed schemes like the transmit-backscatter protocol and transmit-backscatter relay protocol. Additionally, Jia \textit{et al.} proposed a method utilizing Dantzig-Wolfe decomposition and developed column generation-based algorithms for a Low Earth Orbit (LEO) satellite-assisted UAV system. Their approach aimed to minimize energy costs while efficiently managing data transmission for IoRT sensors and optimizing UAV trajectories \cite{9184929}. Similarly, Dong \textit{et al.} used ADS-B information to enhance UAV trajectory planning in urban airspace, introducing SSP and PSO-RRT algorithms to improve flight safety and efficiency \cite{chao2025three}.
Besides, solutions for optimizing total throughput, such as particle swarm optimization (PSO) \cite{10038634} and block coordinate descent (BCD) \cite{9723521}, have been proposed. Nevertheless, these solutions face limitations in terms of system performance due to difficulties in handling non-smooth objective functions while striving to attain the global optimum.

This work enhances achievable rates by integrating caching functionality and backscatter technology. We recognize the non-convex nature of maximizing the total achievable rate over a window time and propose using stochastic approaches based on the genetic algorithm (GA) and PSO. These approaches aim to effectively plan the trajectory and dynamic time-splitting ratio of UAV equipped with backscatter devices, referred to as UB, to optimize the achievable rates of end users. In summary, the following key contributions are listed:
$i)$ we formulate a novel optimization problem for the UAV-integrated wireless communication system under imperfect channel state information (CSI), considering essential factors such as source charging power, UB flight duration, altitude, and transmission power; $ii)$
 we develop two effective meta-heuristic algorithms, namely GA and improved PSO (IPSO), to solve the challenging formulated problem, which helps balance the achievable rate and computational complexity; and $iii)$ numerical results are provided to demonstrate the superior performance of the proposed algorithms in terms of the achievable rate and computation time.
\section{System Model and Problem Statement}

\subsection{System Model}
\begin{figure}[t]
    \centering
\includegraphics[trim=1.5cm 1.5cm 1.5cm 4.5cm, clip=true, width=2.2in]{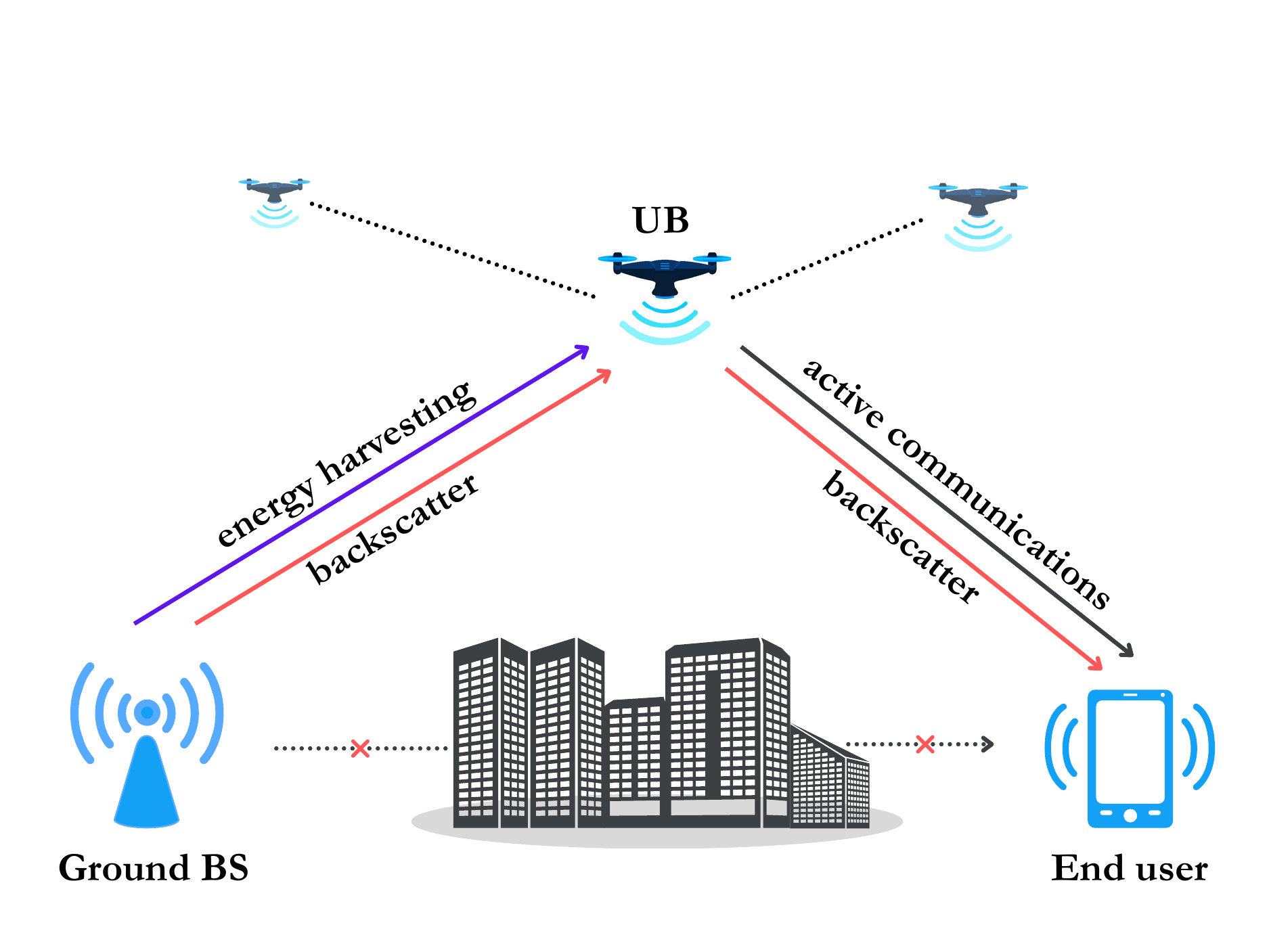}
    \caption{Illustration of the system model, where the GBS serves the end user via a UB.}
    \label{fig:system}
\end{figure}

In challenging communication characterized by significant physical barriers, direct transmission channels from the source to the end user often encounter obstacles. This communication model prioritizes connections between the ground BS (GBS), UB, and the end user, as depicted in Fig.~\ref{fig:system}. The UB is a specialized UAV equipped with a backscatter device. It plays a crucial role in receiving energy from a power source to maintain altitude, movement, and efficient data transmission. The backscattered signal is transmitted from GBS to the end user through the UB, which operates with low energy consumption. Furthermore, the UB features a cache where a subset of the destination's requested information is stored, facilitating proactive communication with the destination.

Assuming that the total flight time of UB is $T$~s and the UB's altitude is fixed at $H$~m. The time interval $T$ is equally divided into $N$ time slots, each with $\sigma_t=T/N$. Therefore, the UB's starting position in time slot $i$ is indicated by $\pmb{\gamma}_i =[x_i,\, y_i,\, z_i]^T$, where $i \in \mathcal{N}\triangleq\{1, \cdots,N\}$. In addition, the source (GBS) and end-user locations are fixed on the ground at $\pmb{\psi}_s = [x_s,\, y_s,\, z_s]^T$ and $\pmb{\psi}_d =[x_d,\, y_d,\, z_d]^T$, respectively. The UB is assumed to fly with a finite maximum speed of $V_{\max}$, resulting in a limited distance in each time slot.

\subsection{Channel Model}
We denote the alternated position in time slot $i$ as $\Delta_i$, with $\pmb{\gamma}_I$ and $\pmb{\gamma}_F$ being the initial and final locations of UB. The movement constraints for UB at time slot $i$ are
\begin{align}
& \|\pmb{\gamma}_{i+1} - \pmb{\gamma}_i\| \triangleq  \Delta_{i} \leq V_{\text{max}} \sigma_t,  i \in \mathcal{N}, \pmb{\gamma}_1 = \pmb{\gamma}_I, \pmb{\gamma}_{N+1} = \pmb{\gamma}_F, \label{eq:1} \\
& d_{tu}^i =  \| \pmb{\gamma}_i - \pmb{\psi}_t\|, \quad t \in \{s, d\}, \quad i \in \mathcal{N}. \label{eq:3}
\end{align} 
We note that \eqref{eq:3} calculates the distance from the source ($s$) to UB  and that from  UB to the end user ($d$) at time slot $i$. Here,  $\pmb{\psi}_t$ represents either the source or the destination location. The channel model encompasses both line-of-sight (LoS) and non-line-of-sight (NLoS) components, thereby incorporating considerations for both large-scale fading and small-scale fading effects \cite{9113440}. The channel estimation between the source/destination (\textit{i.e.}, $t \in \{s, d\}$) and  UB (\textit{i.e.}, $u$) is modeled under the Doppler effect in time slot $i$ as \cite{banh2023uav}
\begin{equation}
  h^i_{tu} = \zeta_i\hat h^i_{tu} + \sqrt{(1-(\zeta_i)^2}\omega^i_{tu},\label{eq:4}
\end{equation}
where $\omega^i_{tu} \sim \mathcal{CN}(0, 1)$ is the channel estimation error, and $\zeta_i$ and $\hat h^i_{tu}$ are the effect of the Doppler effect and the channel coefficient following  \cite{9113440}, \cite{banh2023uav} as
\begin{align}
    &\zeta_i = J^2_0(2\pi f_D^i T_b), \quad \hat{h}^i_{tu} = \sqrt{\beta^i_{tu}} \tilde{h}^i_{tu} = \sqrt{\omega_0 (d^i_{tu})^{-\alpha}} \tilde{h}^i_{tu}, \label{eq:7}
\end{align}
where $J_0(\cdot)$ is the Bessel function of the first kind of order zero; $f_D^i = V_i f_0/c$  is the Doppler frequency with the carrier frequency $f_0$, the speed of light $c$ and the UB's velocity $V_i$; and $T_b$ is the sampling time. In addition, $\beta^i_{tu}$ and $\tilde{h}^i_{tu}$ in \eqref{eq:7} represent large-scale fading and small-scale fading in time slot $i$, respectively. $\omega_0$ is the reference channel gain at $d^i_{tu} = 1$~m, and $\alpha$ denoting the path loss exponent. The small-scale fading $\tilde{h}^i_{tu}$ can be computed as \cite{8901136} with $E\{|\tilde{h}^i_{tu}|^2\} = 1$
\begin{equation}
    \tilde{h}^i_{tu} = \sqrt{\frac{G}{1+G}}\overline{h}^i_{tu}+\sqrt{{\frac{1}{1+G}}} \ddot{h}^i_{tu},\label{eq:9}
\end{equation}
where $\overline{h}^i_{tu}$ accounts for the LoS channel component, $\ddot{h}^i_{tu} \sim \mathcal{CN}(0, 1)$ denotes the Rayleigh fading channel accounting for the NLoS component, and $G$ is the Rician factor.

 \vspace{-0.25cm}
\subsection{Energy Consumption and Energy Harvesting Constraints}
We consider a dynamic time-splitting mechanism where each coherence time can be divided into two dynamic phases. Specifically, a fraction $\delta_i$ and $(1 - \delta_i)$ of the duration $\sigma_t$ are allocated for the backscattering signal and energy harvesting at UB, respectively. During $(1 -\delta_i)\sigma_t$ time, the harvested energy at UB is denoted as \cite{9723521} 
\begin{align}
E^i_h = \sum\nolimits_{i \in \mathcal{N}}
\omega_0 \mu(1 - \delta_i)\sigma_t P_{\text{WPT}} 
/{({d^i_{su}})^{\alpha}},  \label{eq:134}
\end{align}where $\mu$ is the energy harvesting efficiency. To operate UB, the network must provide energy to the UAV to perform tasks such as flying, backscattering, and transmitting power. The energy consumption during time slot $i$ for backscattering, denoted as $E^i_{\text{back}}=\delta_i \sigma_t P_b$, and transmitting power through cache memory, denoted as $E^i_{\text{cach}}=\delta_i \sigma_t P_u$, are defined accordingly. Here, $P_b$ is the UB's circuit power during the backscatter period and $P_u$ is the UB's transmit power. We denote the transmitted power from the GBS to UB as $P_{\text{WPT}}$. The energy consumption for the UB flying  during time slot $i$ is $E^i_{\text{fly}}$ = $\sigma_t P^i_{\text{fly}}$, with $P^i_{\text{fly}}$ denoting the propulsion power consumption as given in \cite{8663615}. Mathematically, $E^i_{\text{fly}}$ can be expressed as
\begin{align}
E^i_{\text{fly}} = \sigma_t\big(P_0 \big(1 + \lambda_1 v^2\big) + 
 P_1 \sqrt{\sqrt{1 + \lambda^2_2 v^4} - \lambda_2 v^2} +  \lambda_3 v^3 \big),  \label{eq:13}
\end{align} where $P_0 \triangleq \frac{\delta}{8} \rho s A \Omega^3 R^3$, $P_1 \triangleq (1 + I) \frac{W^{3/2}}{\sqrt{2\rho A}}$, $\lambda_1 \triangleq \frac{3\sigma_t}{\Omega^2 R^2}$, $\lambda_2 \triangleq \frac{1}{2v^2_0}$, $\lambda_3 \triangleq 0.5a_0 \rho s A$, and $v\triangleq\frac{\Delta_i}{\sigma_t}$ with $\Delta_i$ given in \eqref{eq:1}. More detailed explanations about the parameters can be found in Table I of \cite{8663615}. Note that blade profile power, parasite power, and induced power are required to overcome the profile drag of the blades, the fuselage drag, and the induced drag of the blades, respectively. 
Let denote $E^i_{\text{con}}$ as the consumption energy, including the flying energy, backscatter energy, and cache energy, which is computed as $E^i_\text{con} = E^i_\text{fly} + E^i_\text{back} + E^i_\text{cach}$. 
 \vspace{-0.25cm}
\subsection{Achievable Throughput}
Given the aforementioned setup, the achievable rate obtained at the UB and end user can be expressed as 
\begin{align}
& {R}^i_u = B \log_2 \Big(1 + {e^{-E} \omega_0\zeta_i^2 P_s}/{((d^i_{su})^{\alpha} \tilde{\sigma}^i_u)}\Big),
\label{eq:14} \\
& {R}^i_d = B \log_2 \Big(1 + \frac{e^{-E}\omega_0 \left(\zeta_i^4 \eta^i \omega_0 P_s + \zeta_i^2 \bar{P}_u (d^i_{su})^{\alpha}\right)}{(d^i_{su} d^i_{du})^{\alpha}\tilde{\sigma}^i_d}\Big),
\label{eq:15}
\end{align}
where $B$ is the system bandwidth, ${\tilde{\sigma}^i}_u \triangleq \sigma^2_u + (1 - \zeta_i^2)\sigma^2_n$, ${\tilde{\sigma}^i}_d \triangleq \sigma^2_d + (1 - \zeta_i^2)\sigma^2_n + (1 - \zeta_i^2)^2\sigma^4_n$, $\eta^i$ is the backscatter coefficient during time slot $i$, and $\sigma_u, \sigma_d, \sigma_n$ are the variance of additive white Gaussian noise (AWGN). $E$ is  the Euler-Mascheroni constant, i.e., $E = 0.5772156649$ \cite{9723521} and $\bar{P}_u$ is $\lceil \tau \rceil P_u$ with $\tau$ being a fraction of the requested data stored in the UB. $P_s$ is the source's transmit power.
 \vspace{-0.25cm}
\subsection{Problem Formulation}
The main objective is to maximize the achievable rate received at the end user by jointly optimizing the UB trajectory $\pmb{\gamma} \triangleq \{\pmb{\gamma}_i\}_{i\in\mathcal{N}}$ and the dynamic time-splitting $\pmb{\delta} \triangleq \{\delta_i\}_{ i\in\mathcal{N}}$. Considering practical constraints, the optimization problem is 
\begin{subequations}
\begin{align}
(\mathcal{P}_1): \underset{\pmb{\delta}, \pmb{\gamma}}{\max}
&\quad \sum\nolimits_{i \in \mathcal{N}} R^i_d \label{eq:19}\\
\text{subject to} &\, \, \tau \xi + \sum\nolimits_{i \in \mathcal{N}} R^i_u  \geq \sum\nolimits_{i \in \mathcal{N}} R^i_d, \label{eq:20}\\
&\,\sum\nolimits_{i \in \mathcal{N}} R^i_d \geq \xi, \label{eq:21}\\
&\,\sum\nolimits_{i \in \mathcal{N}} E^i_{\text{fly}} + E^i_\text{back} + E^i_\text{cach} \leq \sum\nolimits_{i \in \mathcal{N}} E^i_h,\label{eq:22}\\
&\,\lVert \pmb{\gamma}_{i+1} - \pmb{\gamma}_i \rVert  \leq V_{\text{max}} \sigma_t, \, \forall i \in \mathcal{N}, \label{eq:23}\\
&\pmb{\gamma}_1 = \pmb{\gamma}_I, \, \pmb{\gamma}_{N+1} = \pmb{\gamma}_F,\, 0 \leq \delta_i \leq 1,  \forall i \in \mathcal{N}, \label{eq:25}
\end{align}
\end{subequations}
where $\xi$ (in bps) represents the data rate requested by the end user. In problem $\mathcal{P}_1$,  constraint \eqref{eq:20} guarantees 
efficient data transfer and storage management. Constraint \eqref{eq:21} ensures that the achievable rate received matches the user's request. Constraint \eqref{eq:22} is imposed to ensure the UB’s sustained operation by requiring sufficient total harvested energy for signal transmission and movement. Constraint \eqref{eq:23} imposes limits on the UB's speed to prevent it from exceeding operational capabilities. The designated trajectory of UB and the dynamic time-splitting ratio is specified in constraint \eqref{eq:25}.

 \vspace{-0.25cm}
\section{Genetic Algorithm}
GA is well-suited for solving complex, non-convex optimization problems where obtaining exact solutions is computationally expensive or impractical. By performing parallel searches, preserving population diversity, and employing probabilistic selection rules, it enhances the likelihood of finding high-quality solutions as outlined in Algorithm~\ref{pseudo:GA}.

\noindent\textbf{Solution encoding:} Each solution is represented by an individual through a genetic form. The optimization variables ($x_i, y_i, z_i$)  are normalized to the range of $[0, 1]$ to facilitate subsequent calculations with $\upsilon \in \{x,y,z\}$ as follows:
\begin{equation}
\upsilon_{\text{normalize}} = (\upsilon_{\text{real}} - \upsilon_{\text{min}})/(\upsilon_{\text{max}} - \upsilon_{\text{min}}).
\label{for3.2}
\end{equation}

\noindent\textbf{Heuristic initialization:}
The UB's coordinates are initialized by a normal distribution $\upsilon \sim \mathcal{N}(\upsilon_{\text{mean}}, \sigma_\upsilon^2)$. One can adjust $\sigma_\upsilon^2$ to diversify the solution within the considered duration.

\noindent\textbf{Fitness function:} If a potential solution violates at least one constraint in problem $\mathcal{P}_1$, the fitness value is assigned a small fixed number. Otherwise, the fitness is computed as $-\sum_{i=1}^{N} R_{d}^{i}$:
\begin{equation}
\mathrm{Fitness} = 
    \begin{cases} 
    -1, & \text{if a constraint is violated}, \\
    - \sum_{i=1}^{N} R^{i}_{d}, & \text{otherwise.}
    \end{cases}
\label{eq:fitness} 
\end{equation}

\noindent\textbf{Selection:}
The selection process employs two principal strategies: elitism, where a subset of the fittest individuals are directly chosen for reproduction and the roulette wheel selection for choosing the remaining individuals. Specifically, in a population comprising $n$ individuals, each individual $k$ with a fitness value $\mathrm{Fitness}(k)$ defined similarly as in \eqref{eq:fitness}, the probability of selection is proportionate to these fitness values as
\begin{equation}
p_k = \mathrm{Fitness}(k)\big/{\sum\nolimits_{j=1}^{n} \mathrm{Fitness}(j)}.
\end{equation}

\noindent\textbf{Crossover:}
The process results in two corresponding progenies. The probability of a gene undergoing crossover is denoted by $\epsilon_c$. The genetic constitution of the offspring is influenced by the numeric factor $\phi$. The value of the gene $i$ for both offspring is computed as
\begin{equation}
c_{i}^{1} = \phi_{i}pa_{i} + (1 - \phi_{i})pa'_{i}, \hspace{0.5cm} c_{i}^{2} = \phi_{i}pa'_{i} + (1 - \phi_{i})pa_{i},
\label{eq:c2}
\end{equation}
where $\phi_{i}$ is uniformly selected in the range $(0, 1)$. Furthermore, $pa_{i}$ and $pa'_{i}$ denote the $i$-th gene value of the respective parents selected for the crossover.

\noindent\textbf{Mutation:} 
Through Gaussian mutation, each gene in the recently formed offspring can experience mutation with a probability $\epsilon_m$. The value at the mutation point is incremented by a quantity following a Gaussian distribution:
\begin{equation}
c_{i}' = c_{i} + \mathcal{N}(0, U^{-2}),
\label{eq:mu}
\end{equation}
where $c_{i}'$ and $c_{i}$ are the mutated gene value and the original gene value, respectively, while $U$ is a parameter that affects the standard deviation of the Gaussian distribution.

\noindent\textbf{Individual adjustment:}
In the process of crossover and mutation in GA, a small increase is added to the genes. However, this process may cause deviations from the desired range of gene values. Consequently, a correction mechanism is implemented to ensure that GA maintains a good solution.

\begin{algorithm}[t]
    \caption{GA-based Algorithm for Solving Problem $\mathcal{P}_1$}
    \label{pseudo:GA}
    \begin{algorithmic}[1]
        \renewcommand{\algorithmicrequire}{\textbf{Input:}}
        \renewcommand{\algorithmicensure}{\textbf{Output:}}
        \REQUIRE Population size $S$, maximum number of generations $G$, crossover rate $\epsilon_c$, and mutation rate $\epsilon_m$.
        \ENSURE The best solution to $\delta$ and $\pmb{\gamma}$
        \STATE Initialize population with size $S$;
        \STATE Evaluate the fitness of each individual $k_i$ by \eqref{eq:fitness};
        \FOR{\textbf{all} $i = 1: G$}
            \STATE Arithmetic crossover (AMXO) with a rate $\epsilon_c$ using \eqref{eq:c2};
            \STATE Mutation on the offspring produced by crossover with a rate $\epsilon_m$ using \eqref{eq:mu};
            \STATE Apply individual adjustment to the offspring produced by crossover and mutation;
            \STATE Evaluate the fitness of each individual $k_i$  using \eqref{eq:fitness};
            \STATE Sort the individuals by their fitness and perform the selection of the new population;
        \ENDFOR
        \RETURN The best individual in the population.
    \end{algorithmic}
\end{algorithm}

\textit{Computational complexity:} The computational complexity of GA is $\mathcal{O}\big(S \log(S)G(3N + 3)^2\big)$, where $3N + 3$ is the number of dimensions of a solution space, $S$ is the size of the population, and $G$ indicates the maximum number of generations to reach convergence. Generally, GA exhibits slow convergence.

\section{Improved Particle Swarm Optimization}
In PSO, particles interact with each other, influencing movement directions based on individual changes. 
This paper enhances PSO through mutation and parameter adjustments.

\noindent\textbf{Initialization of the population:}
Each particle in PSO is characterized by its position, velocity, and experience. A particle at position $X^k = [x_k,\, y_k,\, z_k]^T$ can be initialized following $X^k \sim \mathcal{N}({X^k}_{\text{mean}}, \sigma_{X^k}^2)$, with an initial velocity $v^k_0 = 0$ and the best known personal experience $P_{\text{best}}^k = X^k$. The best position of the swarm, $G_{\text{best}}$, is updated as
\begin{equation}
G_{\text{best}} = \underset{k \in P}{\mathrm{argmin}} \; \mathrm{Fitness}(k).
\label{eq:233}
\end{equation}

\noindent\textbf{Velocity update:}
The velocity is affected by two components, $P_{\text{best}}$ and $G_{\text{best}}$. It is calculated using random numbers to adjust the velocity towards the positions of $P_{\text{best}}$ and $G_{\text{best}}$. The velocity update equation for each particle is expressed as
\begin{equation}
v_{i+1}^k = wv_i^k + c_1r_1(P_{\text{best}}^k - X_i^k) + c_2r_2(G_{\text{best}} - X_i^k),
\label{eq:24}
\end{equation}
where $v_{i+1}^k$ and $v_i^k$ are the velocities of particle $k$ at iteration $i+1$ and $i$, respectively, $w$ is the inertia weight, $r_1$ and $r_2$ are random numbers drawn from a uniform distribution within the interval $(0, 1)$, and $c_1$ and $c_2$ are acceleration coefficients that represent cognitive and social components, respectively. Additionally, $X^k_i$ is the particle $X^k$ in $i$-th generation. 

\noindent\textbf{Position update:}
After updating the velocity, the particle $k$ moves to a new position with the updated velocity:
\begin{equation}
X_{i+1}^k = X_i^k + v_{i+1}^k.
\label{eq:255}
\end{equation}

\noindent\textbf{Improvement:}
We introduce two steps, mutation and adjustment of the inertia weight, to enhance the performance, expedite convergence, and expand the search space of the conventional PSO algorithm \cite{alireza2011pso}.

\textit{Mutation:} A Gaussian mutation approach proposed in  \cite{1202250} enhances the exploratory ability of PSO. This approach involves mutating some genes of a particle at position $X^k$  as
\begin{equation}
X_{i}^k = X_{i}^k + \mathcal{N}(0,\; 0.1).
\label{Imutation}
\end{equation}

\textit{Adjustment of the inertia weight:}
The inertia weight parameter~\eqref{eq:24} controls velocity updates in PSO, balancing exploration and exploitation. A large inertia weight encourages global search, while a smaller weight favors local convergence. Kennedy \textit{et al.} \cite{kennedy1998parameter} demonstrated that linearly decreasing the inertia weight improves solution quality by transitioning PSO from global to local search, following:
\begin{equation}
w = w_{\text{max}} - (w_{\text{max}} - w_{\text{min}}) \left( {t}/{T} \right)^{0.5},
\label{eq:277}
\end{equation}
 where $t$ and $T$ denote the current and total iteration numbers, respectively. The term $(t/T)^{0.5}$ ensures effective weight regulation over a large number of generations, optimizing the trade-off between exploration and convergence.

\noindent\textbf{Individual adjustment:}
Similar to the individual adjustment process of GA, a correction mechanism is implemented to ensure that the gene values are within the range of $[0,\, 1]$ after the updated position and mutation process in PSO.

\textit{Computational complexity:} The proposed IPSO is shown in Algorithm \ref{pseudo:IPSO} with
the complexity in the order of \(\mathcal{O}(SG(3N + 3)^2 )\), where \(3N + 3\) is the dimensions of a solution space. 

\begin{algorithm}[t]
    \caption{IPSO Algorithm for Solving Problem $\mathcal{P}_1$}
    \label{pseudo:IPSO}
    \begin{algorithmic}[1]
    \fontsize{9}{9}\selectfont
        \renewcommand{\algorithmicrequire}{\textbf{Input: }}
        \renewcommand{\algorithmicensure}{\textbf{Output: }}
        \REQUIRE Population size $S$, maximum number of iterations $G$, coefficients $c_1, c_2$, and the inertia weight $w_{\max}$ and $ w_{\min}$.
        \ENSURE The best solution to $\delta$ and $\pmb{\gamma}$.
        \FOR{$k = 1$ to $S$}
            \STATE Initialize the particle position $X^k$ and velocity $X^k = 0$;
        \ENDFOR
        \STATE Update the global best $G_{\text{best}}$ according to \eqref{eq:233};
        \FOR{$i = 1$ to $G$}
            \FOR{$k = 1$ to $S$}
                \FOR{$d = 1$ to the number of dimensions}
                    \STATE Initialize random numbers $r_1,\, r_2 \sim U(0,1)$;
                    \STATE Update the velocity of the particle using \eqref{eq:24};
                   \ENDFOR
                \STATE Update new position of particle $X^k_i$ using \eqref{eq:255};   
                \STATE Adjust the particle to standardization;
                \STATE Evaluate the fitness of particle $X^k_i$ using \eqref{eq:fitness};
                \IF{$\mathrm{Fitness}$($X^k_i$) $<$ $\mathrm{Fitness}$($P^k_{\text{best}}$)}
                    \STATE Update $P^k_{\text{best}}$;
                \ENDIF
            \ENDFOR
            \STATE Update the global best $G_{\text{best}}$ using \eqref{eq:233};
            \STATE Perform a mutation on position $X^k_i$ using \eqref{Imutation};
            \STATE Update the inertia weight $w$ using \eqref{eq:277};
        \ENDFOR
        \RETURN The best particle in the population.
    \end{algorithmic}
\end{algorithm}

\vspace{-0.25cm}
\section{Numerical Experiments}
We consider a system model involving communication links between the GBS and UB,  the links between the UB and the end user, with datasets from \cite{9723521} and \cite{8663615}. GA uses a population size of $100$ with a maximum of 6,000 generations and allows up to 200 iterations without quality improvement. The crossover and mutation rates are set at 80\% and 10\%, respectively. The inertia weight is adjusted between 0.1 and 0.9 for IPSO to match GA. The acceleration coefficients for the social and cognitive components are both 1.5. The proposed algorithms are compared with state-of-the-art benchmarks comprising BCD \cite{9723521} and PSO \cite{10038634}. 
\begin{figure}[t]
     \captionsetup{justification=justified,singlelinecheck=false}
     \begin{subfigure}[t]{0.24\textwidth}
         \includegraphics[trim=3.5cm 8.4cm 3.0cm 8.5cm, clip=true, scale=0.3]{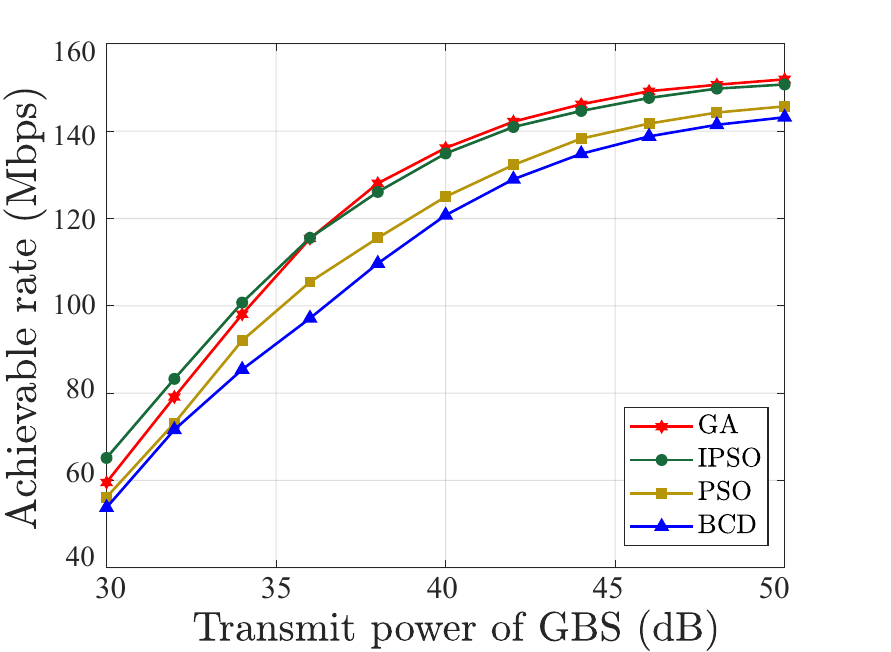}
         \caption{Achievable rate versus $P_{\text{WPT}}$}
         \label{fig:P_WPT}
     \end{subfigure}
     \hfill
     \begin{subfigure}[t]{0.24\textwidth}
         \includegraphics[trim=3.6cm 8.4cm 4.0cm 8.5cm, clip=true, scale=0.3]{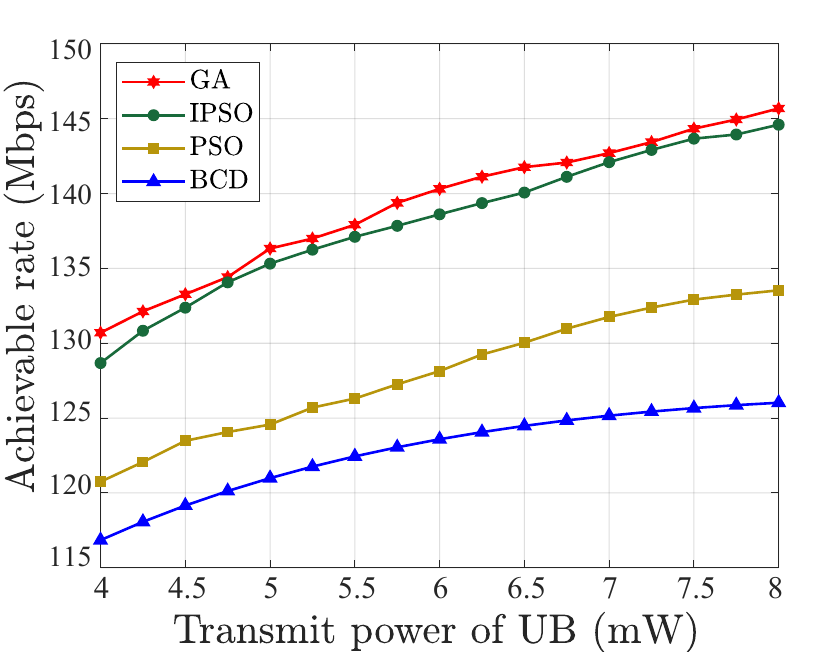}
         \caption{Achievable rate versus $P_{u}$}
         \label{ig:Pu_change}
     \end{subfigure} 
        \caption{Achievable rate versus transmit power of GBS and UB.}
        \label{fig:power_change}
        \vspace{-0.25cm}
\end{figure}

Fig.~\ref{fig:power_change}(a) shows the achievable rate of all four benchmarks increasing almost linearly with the rise in charging power from GBS. Notably, IPSO outperforms GA at lower power levels but trails behind GA as charging power increases. 
Comparatively, PSO yields lower achievable rates than other stochastic algorithms. All stochastic algorithms exhibit superior results compared to BCD  \cite{9723521}. The discrepancy arises because BCD only converges to a Karush-Kuhn-Tucker solution. Meanwhile, the proposed algorithms can exceed local optimal points, toward the optimal solution. Fig.~\ref{fig:power_change}(b) shows the relationship between UB transmission power and achievable data rate. IPSO and GA surpass the remaining benchmarks. 

\begin{figure}[t]
     \captionsetup{justification=justified,singlelinecheck=false}
     \begin{subfigure}[t]{0.24\textwidth}
         \includegraphics[trim=3.5cm 8.4cm 4.0cm 9.0cm, clip=true, scale=0.3]{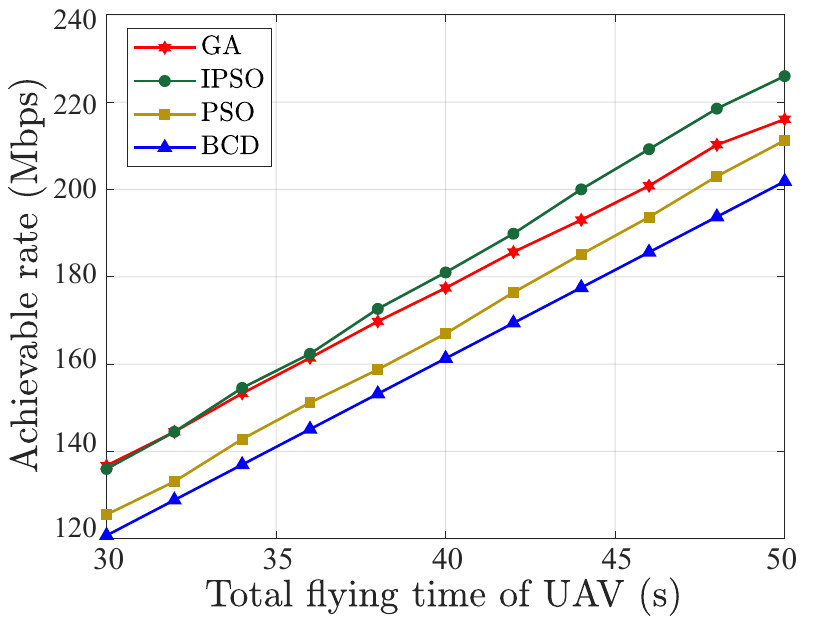}
         \caption{Achievable rate versus $T$}
         \label{fig:Time_change}
     \end{subfigure}
     \hfill
     \begin{subfigure}[t]{0.24\textwidth}
         \includegraphics[trim=3.5cm 8.4cm 4.0cm 9.0cm, clip=true, scale=0.3]{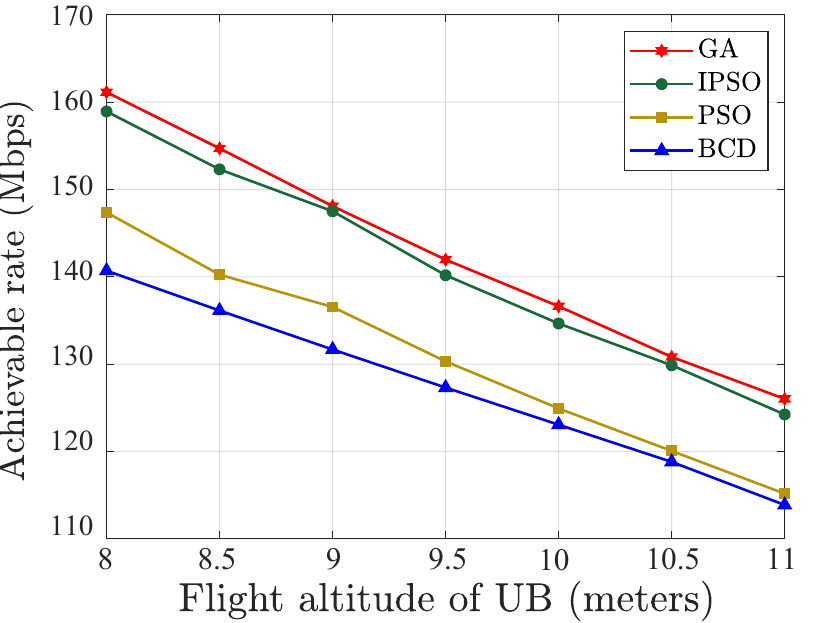}
         \caption{Achievable rate versus $H$}
         \label{fig:H_change}
     \end{subfigure}      
        \caption{Achievable rate v.s. flying time and altitude of UB.}
         \vspace{-0.25cm}
        \label{fig:some_change}
\end{figure}
As the UB's flight time increases, an upward trend in achievable rates is depicted in Fig.~\ref{fig:some_change}(a). Conversely, Fig.~\ref{fig:some_change}(b) depicts a downward trend across all schemes as the flight altitude of the UB increases. With an increase in flight time ($T$), The network's total achievable rate improves significantly. Both GA and IPSO consistently outperform the two baseline methods. However, as altitude increases, GA consistently achieves the highest rates compared to the other benchmarks. 
\begin{figure}[t]
     \captionsetup{justification=justified,singlelinecheck=false}
     \begin{subfigure}[t]{0.24\textwidth}
         \includegraphics[trim=3.5cm 8.0cm 4.0cm 8.5cm, clip=true, scale=0.3]{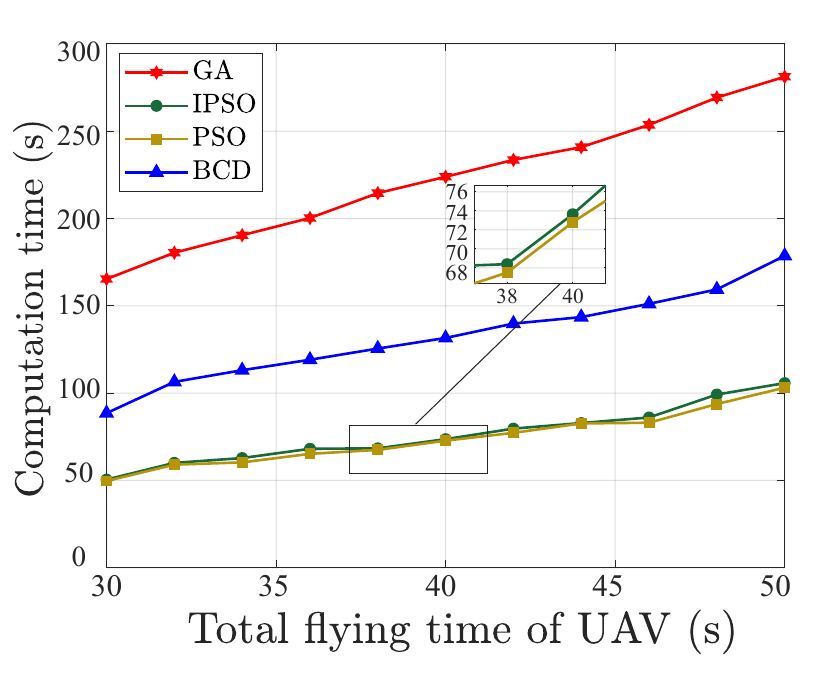}
         \caption{Computation time}
         \label{fig:time_complex}
     \end{subfigure}
     \hfill
     \begin{subfigure}[t]{0.24\textwidth}
         \includegraphics[trim=3.5cm 8.0cm 4.0cm 8.5cm, clip=true, scale=0.3]{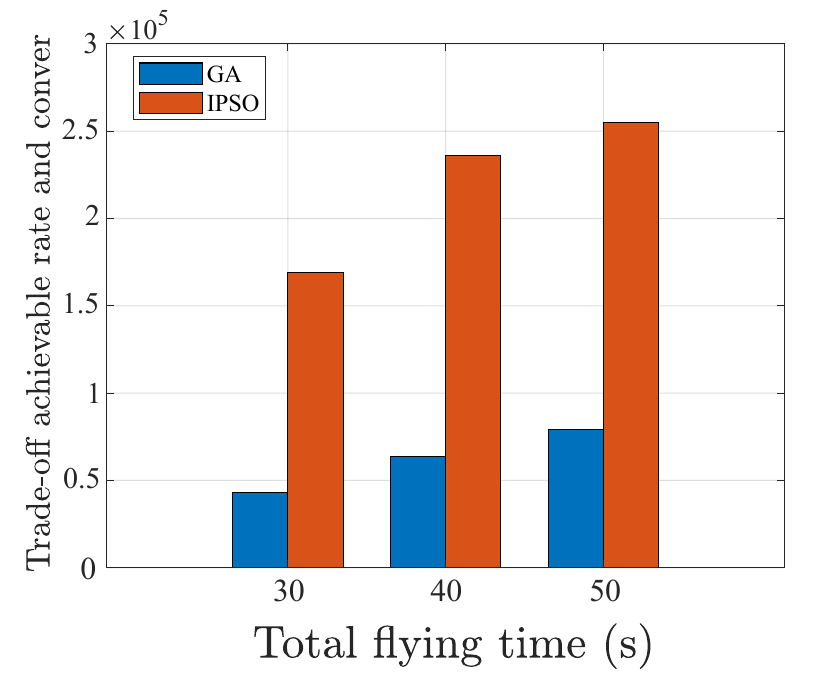}
         \caption{Ratio of  rate and generations}
         \label{fig:convergence}
     \end{subfigure}        
        \caption{Analysis of computation time and complexity.}
        \label{fig:performance}
         \vspace{-0.25cm}
\end{figure}
Fig.~\ref{fig:performance}(a) illustrates the computation time. BCD demonstrates a lower cost than GA, suggesting improved speed. Both PSO and IPSO exhibit relatively low running times.  
 Fig.~\ref{fig:performance}(b) depicts the influence of the number of genes on the convergence of the proposed algorithms across the flight time intervals. We compute the quotient of the achievable rate and the number of generations required for convergence. IPSO shows a noticeable improvement compared to GA.

Figs.~\ref{fig:Trapower_change}(a) and \ref{fig:Trapower_change}(b) illustrate the UAV’s trajectory and dynamic time splitting with $P_\text{WPT} = 30$ dB and $P_\text{WPT} = 40$ dB.  The trend of the UAV  tends to focus on charging power when it is farther from the UE and closer to GBS while prioritizing data transmission when it is near the UE and farther from the GBS. The Gantt chart next to the 3D chart shows the dynamic time-splitting ratio. At the highest point on the chart, represented in yellow, the UAV is fully dedicated to data transmission and focused on charging energy via WPT.
\renewcommand{\figurename}{Fig.}
\renewcommand{\thefigure}{5}
\begin{figure}[t]
     \captionsetup{justification=justified,singlelinecheck=false}
     \begin{subfigure}[t]{0.24\textwidth}
         \includegraphics[trim=9.5cm 0.5cm 5.0cm 0.0cm, clip=true, scale=0.28]{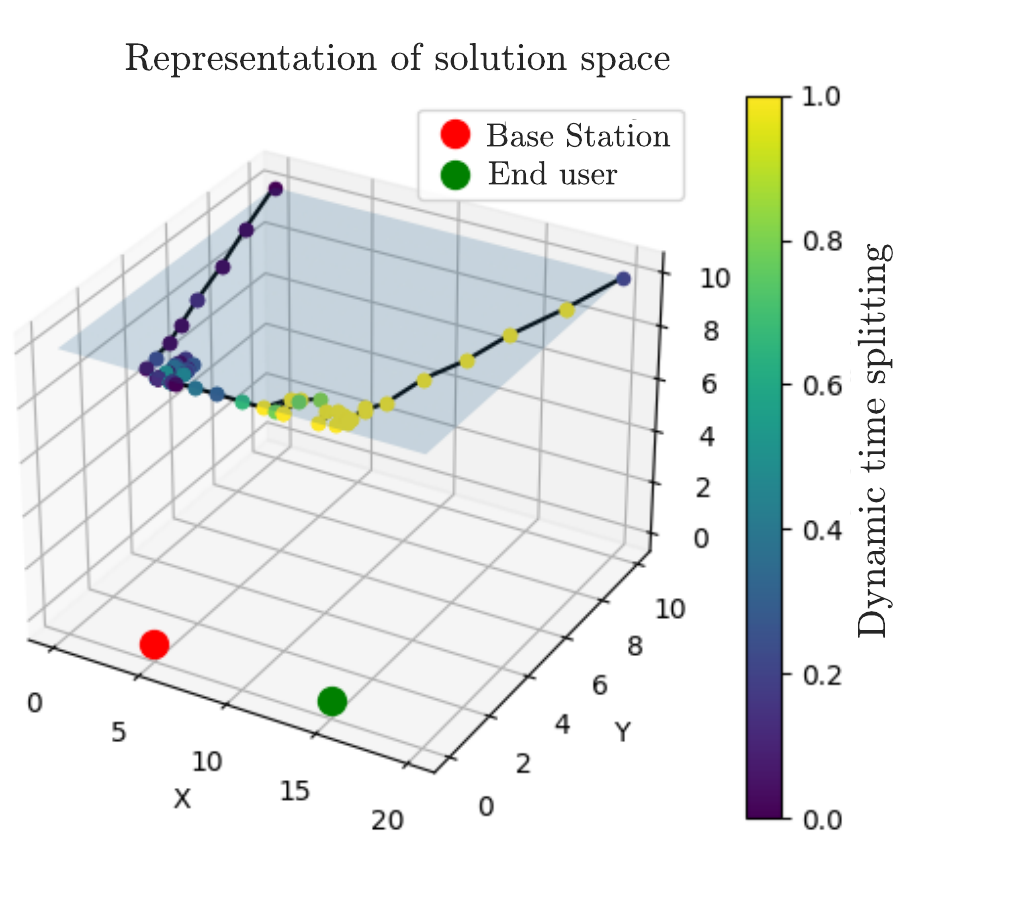}
         \caption{$P_{\text{WPT}} = 30$dB}
         \label{fig:TraP_WPT}
     \end{subfigure}
     \hfill
     \begin{subfigure}[t]{0.24\textwidth}
         \includegraphics[trim=9.5cm 0.5cm 5.0cm 0.0cm, clip=true, scale=0.28]{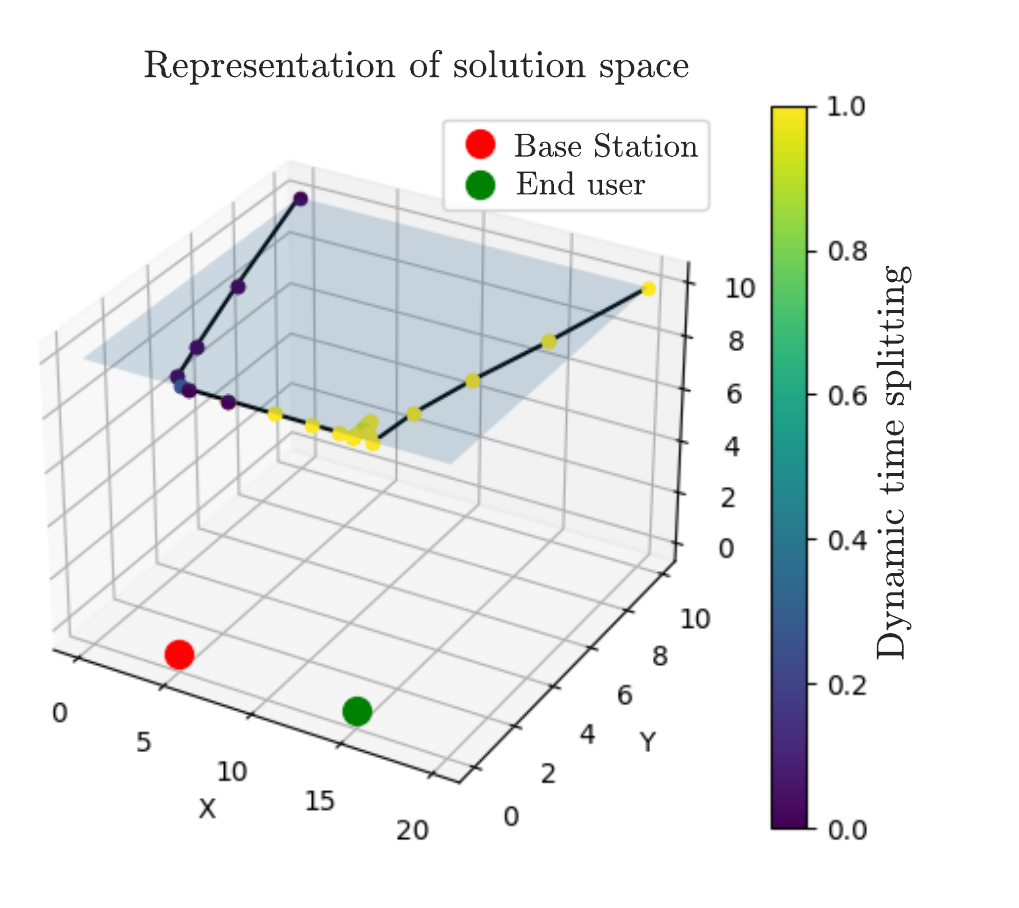}
         \caption{$P_{\text{WPT}} = 40$dB}
         \label{fig:TraPu_change}
     \end{subfigure}        
        \caption{Trajectory of UAV and dynamic time splitting.}
        \label{fig:Trapower_change}
         \vspace{-0.25cm}
\end{figure}

\section{Conclusion}
This paper has investigated the potential for integrating UAVs into radio networks, emphasizing their ability to improve service quality. We have introduced two metaheuristic algorithms to efficiently tackle the nonconvex sum rate maximization problem and demonstrate superior performance. It enriches the ongoing discourse on next-generation communication systems, emphasizing the pivotal role of UAVs in providing high-quality communication services.
 \vspace{-0.25cm}

\bibliographystyle{IEEEtran}
\bibliography{IEEEabrv,refs}
\end{document}